\newif\ifsubmode
\submodefalse

\newif\ifprintfig
\printfigtrue

\newif\ifemulate
\emulatetrue

\ifsubmode
  \documentclass[12pt,preprint]{aastex}
  \received{}
  \accepted{}
  \journalid{}{}
  \articleid{}{}
\else
   \documentclass{emulateapj}
   \submitted{{\it Submitted for publication in ApJL}}
\fi

\shortauthors{Willman, et al.}
\shorttitle{New Milky Way Satellite}

\def\lesssim{\mathrel{\hbox{\rlap{\hbox{\lower4pt\hbox{$\sim$}}}\hbox{$<$}}}}
\def\gtrsim{\mathrel{\hbox{\rlap{\hbox{\lower4pt\hbox{$\sim$}}}\hbox{$>$}}}}

\begin{document}

\title{A New Milky Way Dwarf Galaxy in Ursa Major}

\author{Beth Willman\altaffilmark{1}, Julianne J. Dalcanton\altaffilmark{2,3}, David Martinez-Delgado\altaffilmark{4,5}, Andrew A. West\altaffilmark{2}, Michael R. Blanton\altaffilmark{1}, David W. Hogg\altaffilmark{1}, J.C. Barentine\altaffilmark{6}, Howard J. Brewington\altaffilmark{6}, Michael Harvanek\altaffilmark{6}, S.J. Kleinman\altaffilmark{6}, Jurek Krzesinski\altaffilmark{6,7}, Dan Long\altaffilmark{6}, Eric H. Neilsen, Jr.\altaffilmark{8}, Atsuko Nitta\altaffilmark{6}, Stephanie A. Snedden\altaffilmark{6}}


\altaffiltext{1}{New York University, Center for Cosmology and Particle Physics, Department of Physics,
4 Washington Place, New York, NY 10003, beth.willman@nyu.edu,
mb144@nyu.edu, david.hogg@nyu.edu}
 
\altaffiltext{2}{Department of Astronomy, University of Washington, Box 351580, Seattle, WA, 98195, jd@astro.washington.edu, west@astro.washington.edu}

\altaffiltext{3}{Alfred P. Sloan Research Fellow}

\altaffiltext{4}{Instituto de Astrophysica de Andalucia (CSIC), Granada, Spain, ddelgado@iaa.es}

\altaffiltext{5}{Max-Planck-Institut für Astronomie, Königstuhl 17, D-69117 Heidelberg, Germany}

\altaffiltext{6}{Apache Point Observatory, 2001 Apache Point Rd., Sunspot, NM  88349, jcb@apo.nmsu.edu, hbrewington@apo.nmsu.edu, harvanek@apo.nmsu.edu, sjnk@apo.nmsu.edu, jurek@apo.nmsu.edu, long@apo.nmsu.edu, ank@apo.nmsu.edu, snedden@apo.nmsu.edu}

\altaffiltext{7}{Mt. Suhora Observatory, Cracow Pedagogical University, ul. Podchorazych 2, 30-084, Cracow, Poland}

\altaffiltext{8}{Fermi National Accelerator Laboratory, PO Box 500, Batavia, IL 60510, neilsen@fnal.gov}


\ifsubmode\else
  \ifemulate\else
     \clearpage
  \fi
\fi


\ifsubmode\else
  \ifemulate\else
     \baselineskip=14pt
  \fi
\fi

\begin{abstract}

In this Letter, we report the discovery of a new dwarf satellite to
the Milky Way, located at ($\alpha_{2000},
\delta_{2000}$) $=$ (158.72,51.92) in the constellation of Ursa Major.  
This object was detected as an overdensity of red, resolved stars in
Sloan Digital Sky Survey data.  The color-magnitude diagram of the
Ursa Major dwarf looks remarkably similar to that of Sextans, the
lowest surface brightness Milky Way companion known, but with
approximately an order of magnitude fewer stars.  Deeper follow-up
imaging confirms this object has an old and metal-poor stellar
population and is $\sim$ 100 kpc away. We roughly estimate M$_V \! =$
-6.75 and $r_{1/2} \! =$ 250 pc for this dwarf.  Its luminosity is
several times fainter than the faintest known Milky Way
dwarf. However, its physical size is typical for dSphs. Even though
its absolute magnitude and size are presently quite uncertain, Ursa
Major is likely the lowest luminosity and lowest surface brightness
galaxy yet known.

\end{abstract}


\ifemulate\else
   \clearpage
\fi

\section{Introduction}

A complete census and study of nearby dwarf galaxies is vital to our
global understanding of galaxy formation.  Dwarf galaxies are the most
numerous type of galaxy in the Universe and are thought to be the
``building blocks'' of larger galaxies.  Milky Way (MW) dwarf galaxies
are particularly interesting because they are close enough for HST to
resolve their stellar populations fainter than their main sequence
turnoffs. This enables precise measurements of dwarfs' structural
parameters, metallicities, and detailed star formation histories when
coupled to wide-field ground-based imaging. MW dwarfs are also close
enough for ground-based spectroscopy to measure the metallicities and
velocities of individual stars.

The existence of dwarf galaxies fainter than those known
also holds promise to substantially improve our understanding of the
``substructure problem''.  Cold dark matter models predict more than
an order of magnitude more low mass dark matter halos than the number
of dwarf galaxies observed around galaxies such as our own
\citep{klypin99,moore99}.  The fraction of low-mass halos that may host
a luminous galaxy is reduced by baryonic physics such as reionization,
feedback, and tidal effects.  However, possible incompleteness in the
census of MW dwarf galaxies at the faint end hinders our
interpretation of such models, leaving open the possibility that they
do not produce the true population \citep{willman04}.

To improve the completeness of the known Milky Way dwarf galaxy
population, we have been conducting a search for Milky Way
satellites in the Sloan Digital Sky Survey (SDSS;
\citealt{willman02}). Careful analyses of resolved stars in both SDSS
and 2MASS have already resulted in the discovery of a new Milky Way
companion
\citep{willman05} and a faint M31 dwarf satellite
\citep{zucker04}, as well as large-scale stellar structures and dwarf
galaxy remnants around the Milky Way {\citep{newberg02, yanny03,
ibata03,rochapinto03,rochapinto04,majewski03,martin04}. However, it
has been more than 10 years since the discovery of the ninth Milky Way
dwarf spheroidal galaxy (\citealt{ibata94}; but see evidence in
\citealt{martin04} and \citealt{delgado05} for a probable
new Milky Way dwarf at low latitude).  In this Letter, we report the
discovery of the Ursa Major dwarf (UMa dSph), the tenth dwarf
spheroidal companion to the Milky Way.

\section{Data and Results}

\subsection{Sloan Digital Sky Survey Data}
The Sloan Digital Sky Survey (SDSS; \citealt{york00}), is a
spectroscopic and photometric survey in 5 passbands ($u,g,r,i,z$;
\citealt{fukugita96,gunn98,smith02}), that has thus far imaged
thousands of square degrees of the sky.  Data is reduced with an 
automatic pipeline consisting of: astrometry
\citep{pier03}; source identification, deblending and photometry
\citep{lupton01}; photometricity determination \citep{hogg01};
calibration \citep{fukugita96,smith02}; and spectroscopic data
processing \citep{edr}.

The Ursa Major dSph was found as part of a systematic search for Milky
Way companions.  It was detected as an 8.5$\sigma$ fluctuation in the density of stars at ($\alpha_{2000},
\delta_{2000}$) $\sim$ (158.72,51.92) with $19.0 < r < 20.5$ and having colors consistent with red giant branch stars.  Although our search algorithm does not produce a perfectly gaussian distribution of stellar surface densities, the detection thresholds are carefully set such that few spurious detections are expected. See \citet{willman02} and \citet{willman05b} 
for a detailed discussien of the detection thresholds, as well as of
our automated search technique, detection limits, and the summarized
survey results. The data relevant for this discovery are publicly
available in Data Release 2 of the SDSS (DR2,
\citealt{dr2}).  

Figure 1 shows color-magnitude diagrams (CMDs) created solely with
SDSS data of the Ursa Major dSph (both before after a statistical subtraction
of field stars; left and middle panels) and the Sextans dSph
(right panel). Sextans is an old and metal-poor ([Fe/H] = -2.1 $\pm$ 0.3;
\citealt{lee03}) Milky Way dSph at a distance of 86 kpc. There are a total of 172 stars in the 200 arcmin$^2$
detection area plotted in the left panel, but only 50 remain after
field subtraction. To perform the statistical subtraction, we first
divided the field and source CMDs into discrete color-magnitude bins,
each bin containing the same number of stars in the field CMD.  The
color-magnitude (CM) bins are large enough such that the density of
field stars can vary substantially within a bin.  We thus subtract the
appropriate (area-normalized) number of stars from each CM bin in the
source CMD by: 1) drawing a location from the density distribution of
field stars within that bin, and 2) removing the source star closest
in color and magnitude to that location. Red giant branch
stars are outlined in the middle panel, and the overdensity at $r \sim
20.5, -0.1 < g-r < 0.5$ is a probable horizontal branch.  We overplot
the stellar locus of the Sextans dSph empirically derived from SDSS
data on its own CMD.

A visual comparison of the Sextans and UMa CMDs shows that they are
strikingly similar, including the details of their horizontal branch
morphologies.  The UMa dSph CMD has roughly an order of magnitude
fewer stars than the Sextans CMD, and thus must have a much lower
surface brightness if it truly is an analogous object.  This is
remarkable given that Sextans is the lowest surface brightness Milky
Way dwarf known, having $\mu_V$ = 26.2 mag arcsec$^{-2}$
\citep{mateo98}, and that the lowest surface brightness dwarf currently
known has $\mu_V$ = 26.8 mag arcsec$^{-2}$
\citep{zucker04}.  We overplot the stellar locus of the Sextans dwarf projected to 100 kpc on Ursa Major's CMD in Figure
2, to illustrate the similarity of their stellar populations.  This
similarity suggests that UMa stars have an [Fe/H] that is similar to
that of Sextans stars.

\placefigure{fig:cmdnoisos}

\placefigure{fig:cmdisos}

\subsection{Isaac Newton Telescope Data}
To confirm the Ursa Major dwarf as a Sextans-like Milky Way companion,
we obtained follow-up imaging with the 2.5m wide-field camera on the
Isaac Newton Telescope (INT) on 2005 March 6-8.  Figure 3 shows a CMD
in Harris $B$ and Sloan $r$ of stars in a 23$'$ $\times$ 12$'$ arcmin
field around the center of UMa from a total of 5600 seconds of
exposure time in $B$ and 4800 seconds in $r$.  The DAOPHOT II/ALLSTAR
package \citep{stetson94} was used to obtain the photometry of the
resolved stars. Sources with $chi$ $<$ 2 and $-0.4 < sharp < 0.4$ are
included in Figure 3.  The magnitudes were calibrated by comparison to
SDSS data.  Because SDSS does not resolve Sextans' main sequence
turnoff, we instead overplot the theoretical isochrone of an [Fe/H] =
-1.7, 13 Gyr old population
\citep{girardi04} projected to 100 kpc.  We used the
\citet{smith02} transformations to convert the Girardi isochrone in Sloan filters from $g$ and $r$
to $B$ and $r$.  The theoretical [Fe/H] of -1.7 is within the
uncertainty (0.3 dex) and intrinsic spread (0.2 dex, Lee et al. 2003)
in the [Fe/H] of Sextans' stars.  Although the close match of the
theoretical isochrone to the UMa data again suggests that the [Fe/H]
of the Ursa Major dwarf is $\sim -2$, the present analysis is not
sufficient to determine its [Fe/H] more precisely.  In addition to the
sparse horizontal and red giant branch seen in the SDSS, a narrow
sub-giant branch becomes clear in these deeper data at 21.5 $< B <$
23.0 and $B-r$ $\sim$ 0.9, confirming UMa as a new MW companion.  A
main-sequence turnoff (MSTO) also appears near $B \sim 24.5$ and $B-r
\sim$ 0.5.  The horizontal branch and MSTO are separated by almost 4
magnitudes in B, characteristic of an old stellar population.  A
distance to UMa of $\sim$ 100 kpc is necessary to match the apparent
magnitudes of the horizontal branch and MSTO stars to those of an old,
metal-poor stellar population.  A detailed analysis of a substantial
amount of data obtained at the INT and other telescopes will be
presented in a subsequent paper.

\placefigure{fig:intcmd}

\subsection{$r_{1/2}$ and $M_V$}
The spatial distribution of red giant branch (RGB) stars outlined in
Figure 1 supports the idea that UMa is a new nearby dwarf.  Its RGB
stellar distribution, shown in the left panel of Figure 4, is very
similar in angular extent to the spatial distribution of Sextans' RGB
stars, shown in the right panel of Figure 4.  Based on this
distribution, the half-light radius of Ursa Major is $\sim$7.75$'$
($r_{\mathrm {1/2}} \!\sim$250 pc, assuming a distance of 100 kpc). This
estimated half-light radius is very similar to that of Sextans, which
has $r_{\mathrm 1/2} \sim$ 200 pc \citep{mateo98}.

\placefigure{fig:distrib}

To estimate the absolute magnitude of UMa, we first sum the
luminosities of stars in Figure 2 (assuming a distance of 100 kpc) to
estimate its faintest possible absolute magnitude $M_{\mathrm
{V,faint}} \! = \! -4.6$. We then apply an approximate correction to
this minimum luminosity by multiplying it by 2 to account for object
stars outside the 200 arcmin$^2$ detection area and then multiplying
by 3 to account for stars that fall below the SDSS magnitude limit.
These multiplicative factors are uncertain and were estimated by
measuring the fraction of light coming from stars brighter than the
horizontal branch in Sextans and Palomar 5 as observed by SDSS.  This
approximation yields $M_{\mathrm {V,corr}}$ = -6.5.  Similarly, we
compare the number of stars in UMa's RGB to the number of stars in
Sextans'.  UMa has $\sim 10$ stars brighter than its horizontal branch
after field subtraction, whereas Sextans has $\sim 100$.  From this
comparison, we estimate $M_{\mathrm {V}}$ = -7.0, because Sextans has
$M_{\mathrm {V}}$ = -9.5.  This size and absolute magnitude is
extremely uncertain, and are only intended to give a sense of UMa's
possible properties.

\section{Discussion}

The absolute magnitude of the UMa dSph is fainter than those of the
faintest known dwarfs but is similar to those of some known globular
clusters.  However, five of the nine known MW dSphs have absolute
magnitudes that also overlap with those of globular clusters. The
half-light size of the Ursa Major dwarf is also quite similar to those of the known Milky Way
dSphs, but is nearly an order of magnitude larger than those of either
the largest Milky Way globular clusters
\citep{harrisGCcat} or the newly discovered extended star clusters around
M31 \citep{huxor04}.  UMa is also more distant than all but one of the
MW globular clusters.  We thus conclude that UMa is a new Milky Way
dwarf spheroidal galaxy.

There is no clear connection between the UMa dwarf and any known
object. The new galaxy is near $(l, b) = (160,54)$, which is not
proximate to any of the known MW dwarfs or globular clusters. UMa does
appear to be located along the great circle possibly traced by the
Magellanic stream, but is more distant (see \citealt{palma02}).  UMa is also
coincidentally located only a few degrees away from SDSS J1049+5103,
another recently discovered Milky Way companion \citep{willman05}.
However, it is nearly a factor of two times farther away.

UMa was detected very close to our detection limits. Numerous other
dwarfs with properties similar to or fainter than the Ursa Major dSph
may thus exist around the Milky Way.  Although no reliable
extrapolation can be made from a single object, the fact that at least
one new Milky Way dwarf was detected in $\sim$ 4700 deg$^2$ ($< 1/8$)
of the sky suggests it is reasonable to expect that 8-9 additional
dwarfs brighter than our detection limits still remain undiscovered
over the entire sky.  If true, that number would preclude models that
do not predict the presence of many ultra-faint dwarfs.  However, our
survey only included sky at $|b| > 30$.  In a scenario where Milky Way
dwarfs are intrinsically biased to lie at high latitude \citep{zentner05}, we would
extrapolate a smaller total number of dwarfs based on this single
detection.

We are in the process of obtaining and analyzing deep, wide-field
imaging of UMa. With these deeper data, we will obtain estimates of
its age and metallicity, as well as measure its detailed spatial
distribution to derive its surface brightness, scale size, and search
for tidal features. 


\acknowledgements
BW and JJD were partially supported by the Alfred P. Sloan Foundation.
MRB and DWH were partially supported by NASA (grant NAG5-11669) and
NSF (grants PHY-0101738 and AST-0428465).  We thank P. B. Stetson,
D. Schlegel, and D. Finkbeiner for helpful conversations and software.
We thank Helio Rocha-Pinto, the referee, for comments that improved
the paper.

The SDSS is managed by the Astrophysical Research Consortium (ARC) for
the Participating Institutions. The Participating Institutions are The
University of Chicago, The U.S. Department of Energy's Fermi National
Accelerator Laboratory, The Institute for Advanced Study, The Japan
Participation Group, The Johns Hopkins University, The Korean
Scientist Group, Los Alamos National Laboratory, the
Max-Planck-Institute for Astronomy (MPIA), the Max-Planck-Institute
for Astrophysics (MPA), New Mexico State University, University of
Pittsburgh, Princeton University, the United States Naval Observatory
and the University of Washington.
  
Funding for the project has been provided by the Alfred P. Sloan
Foundation, the Participating Institutions, the National Aeronautics
and Space Administration, the National Science Foundation, the
U.S. Department of Energy, the Japanese Monbukagakusho, and the Max
Planck Society.


\ifsubmode\else
\baselineskip=10pt
\fi



\clearpage

\begin{figure} 
\plotone{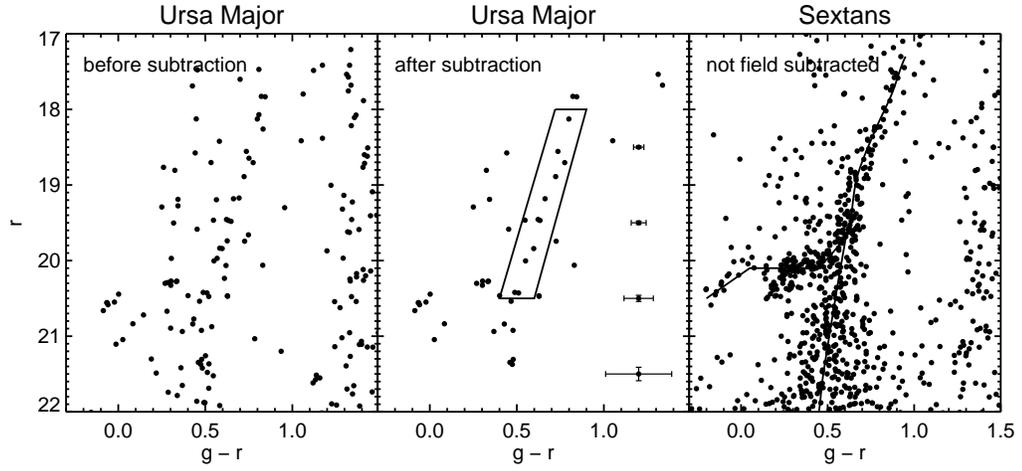}
\caption{ {\it Left Panel:} Ursa Major CMD including all
   172 stars within the 200 arcmin$^2$ area included in the detection,
   without a statistical subtraction of foreground stars. {\it Middle
   Panel:} Field subtracted CMD of UMa.  The probable red giant branch
   of UMa is outlined. {\it Right Panel:} The color-magnitude diagram
   of the Sextans dSph ($\mu_V \!= 26.2$, $d\! = 86$ kpc) without any
   field star subtraction. This CMD includes all stars within Sextans'
   half-light radius and was created solely with SDSS data.  The
   stellar locus of Sextans, empirically measured with the SDSS data,
   is overplotted.  All three CMDs and the field subtraction were
   created solely with SDSS data.}
\label{fig:cmdnoisos} 
\end{figure}

\begin{figure}
\plotone{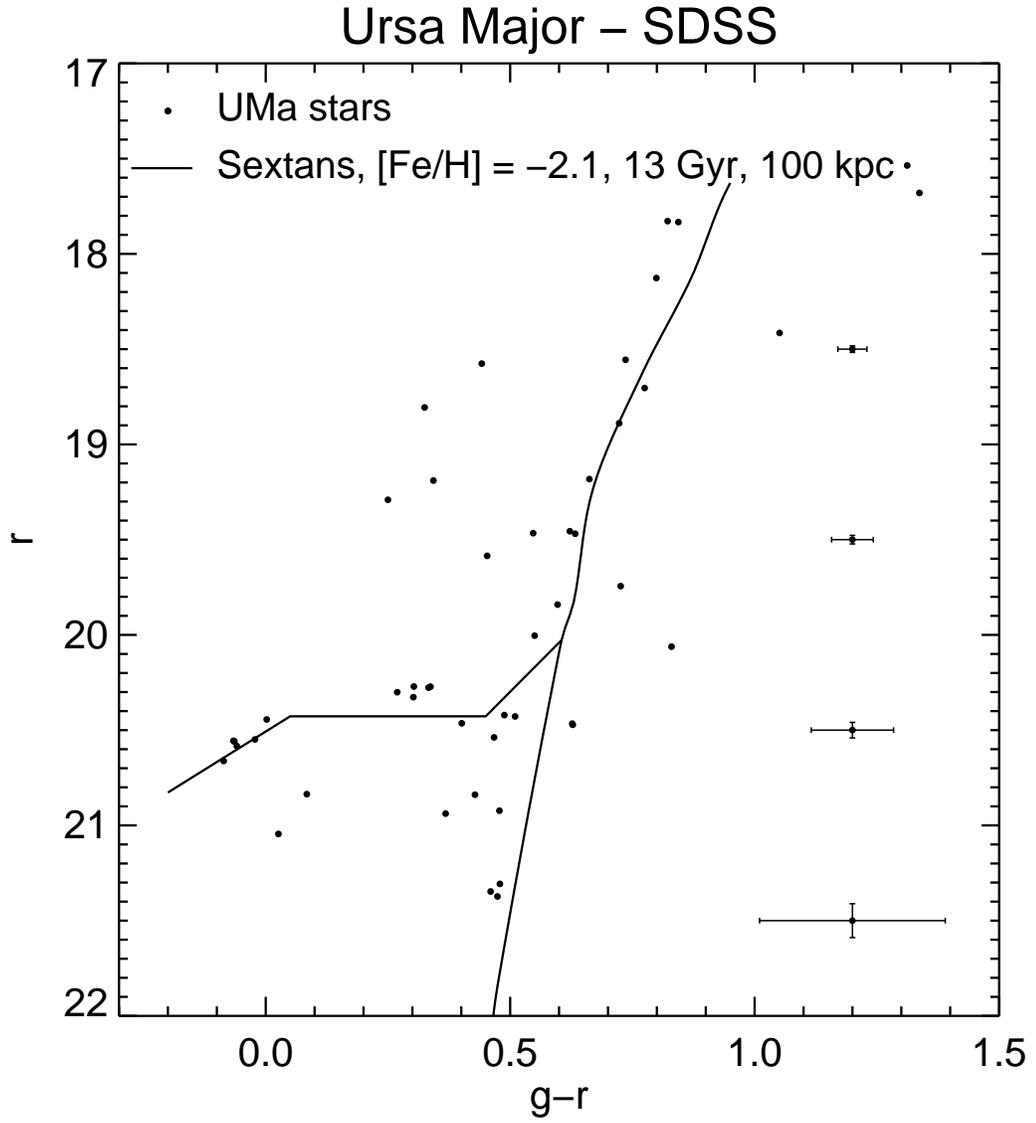}
\caption{The field subtracted color magnitude diagram of Ursa Major. The stellar locus of Sextans stars, empirically
   measured with SDSS data and projected to 100 kpc is overplotted.
   Typical color errors as a function of magnitude are shown at the
   right.}
\label{fig:cmdisos} 
\end{figure}

\begin{figure}
\plotone{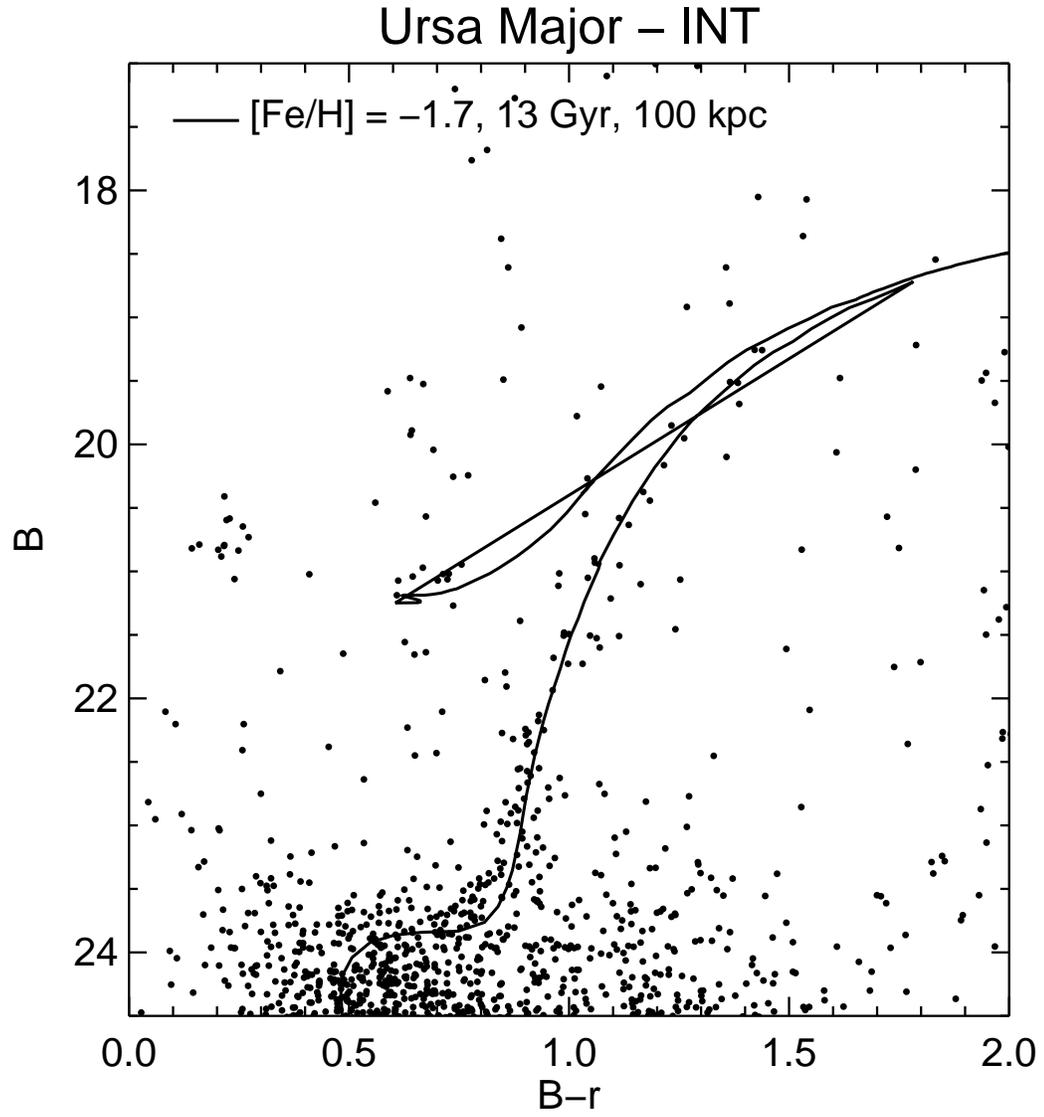}
\caption{The CMD of stars in an 23$'$ $\times$ 12$'$ field around the center of Ursa Major, as observed in a total of 5600 seconds of exposure time in B and 4800 seconds in r.  A theoretical isochrone of an [Fe/H] = -1.7, 13 Gyr old population is overplotted \citep{girardi04}.}
\label{fig:intcmd} 
\end{figure}

\begin{figure}
\plottwo{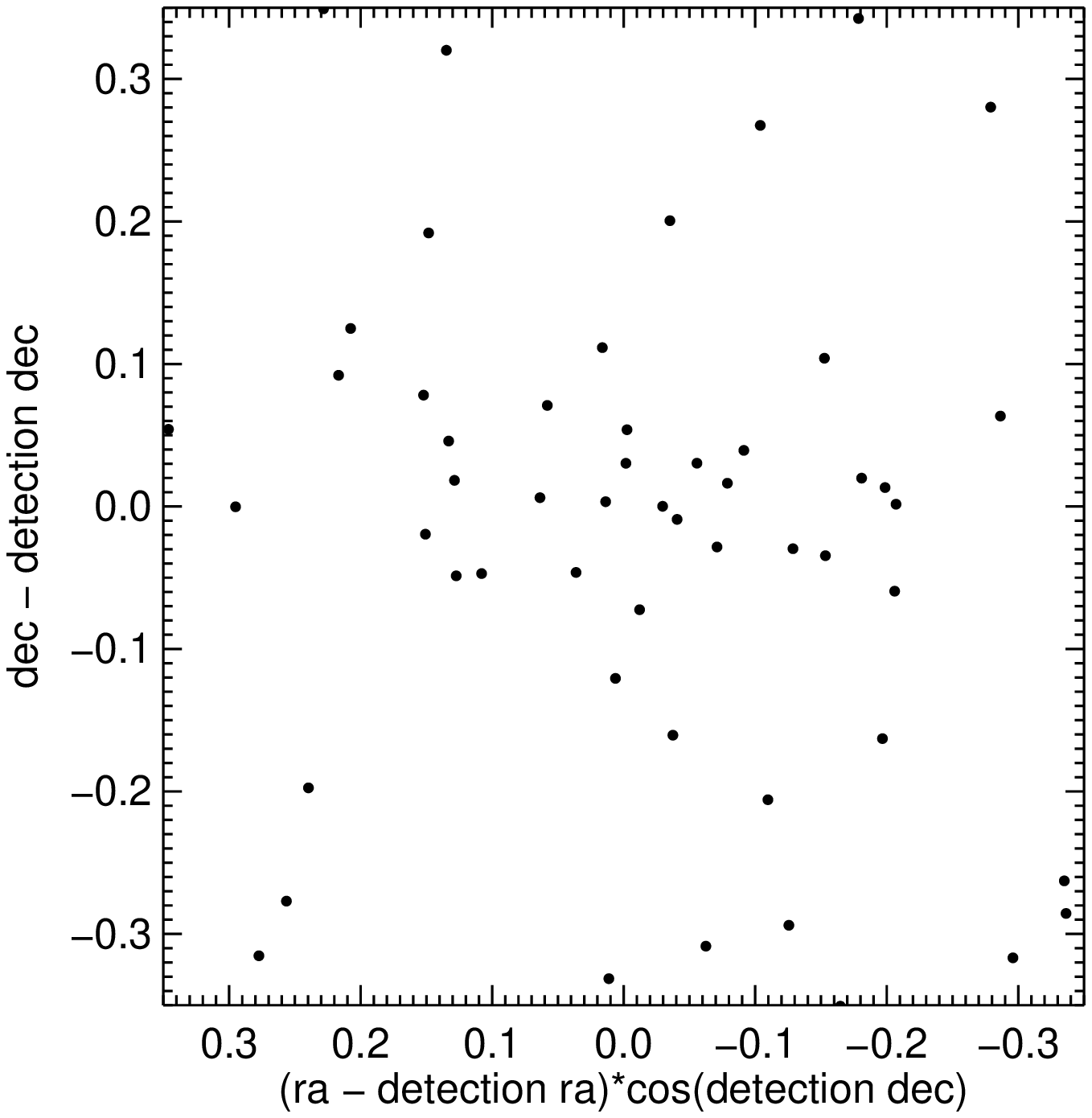}{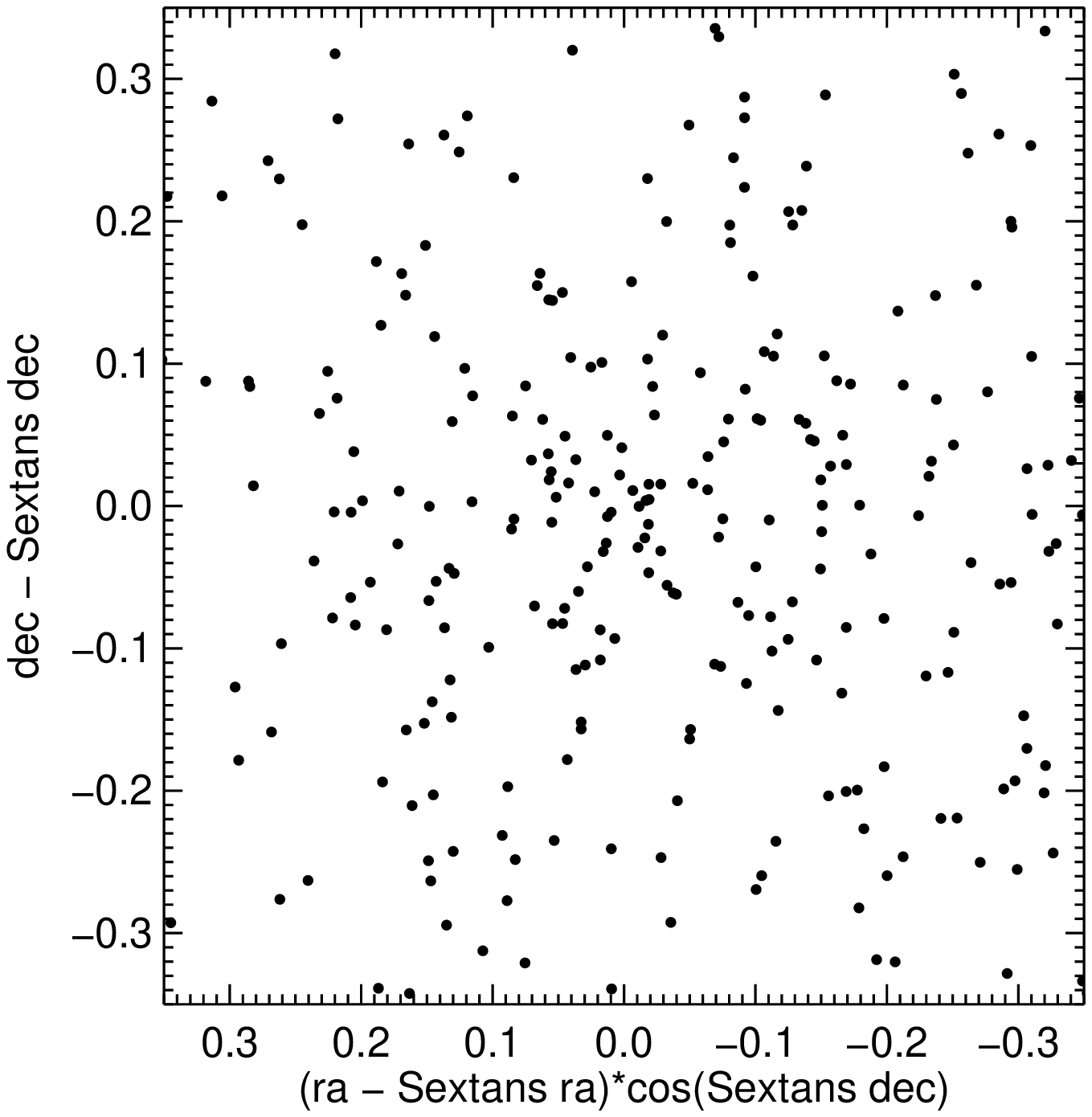}
\caption{{\it Left Panel:} The spatial distribution of Ursa Major as traced by the red
 giant branch stars outlined in the middle panel of Figure 1.  The
 stellar overdensity appears to extend over nearly 0.25 square
 degrees, and its half-light radius is approximately 7.75$'$. {\it
 Right Panel:} For comparison, the spatial distribution of stars in
 the Sextans dSph (d = 86 kpc) that fall in the same region of the
 color-magnitude diagram.}
\label{fig:distrib}
\end{figure}


\clearpage


\ifsubmode\pagestyle{empty}\fi

\end{document}